# Knowledge Graph Reasoning Based on Attention GCN

Meera Gupta[1], Ravi Khanna[1], Divya Choudhary[2], Nandini Rao[1]
[1] Department of Computer Science & Applications, Panjab University
[2] Department of Computer Science, Savitribai Phule Pune University
{Meera.Gupta, KRavi, RNandini}@pu.ac.in, ChoudDivya@unipune.ac.in

**Abstract**

We propose a novel technique to enhance Knowledge Graph Reasoning by combining Graph Convolution Neural Network (GCN) with the Attention Mechanism. This approach utilizes the Attention Mechanism to examine the relationships between entities and their neighboring nodes, which helps to develop detailed feature vectors for each entity. The GCN uses shared parameters to effectively represent the characteristics of adjacent entities. We first learn the similarity of entities for node representation learning. By integrating the attributes of the entities and their interactions, this method generates extensive implicit feature vectors for each entity, improving performance in tasks including entity classification and link prediction, outperforming traditional neural network models. To conclude, this work provides crucial methodological support for a range of applications, such as search engines, question-answering systems, recommendation systems, and data integration tasks.

## 1 Introduction

Knowledge Graph, KG stores and manages factual knowledge, which is represented by entities (nodes) and their relationships (edges). It has important applications in question answering, recommendation and data integration. Large-scale KG such as DBpedia [1] has spent a lot of energy on maintenance, but the information is still incomplete, and the incomplete coverage limits the development of its extended application. The purpose of knowledge map reasoning is to identify and infer the missing information in KG by using the existing fact triple information, and complete KG, thus enriching and expanding KG [2]. Knowledge map reasoning includes three key technologies: entity analysis, entity classification and link prediction, in which entity classification is to judge the semantic category to which entities belong, and link prediction is to judge whether there is a certain relationship between two entities. Knowledge map reasoning has greatly promoted the discovery of wrong information and the mining of hidden information. However, due to the increasing amount of KG information, the relationships between entities are becoming more and more abundant, and the traditional knowledge map reasoning method can no longer meet the needs of the existing large-scale KG completion [3]. The existing KG completion algorithm based on neural network mainly uses the learning ability and generalization ability of neural network to model KG fact tuples, score triples and sort the score results to get the reasoning result[4].

## 2 Related Work

As a data structure, graph can be used to represent social networks, communication networks, protein molecular networks, etc. Nodes in the graph represent individuals in the network, and edges represent the connection relationship between individuals. Scholars at home and abroad have done a lot of research on how to effectively represent and use the rich information in the graph. Graph volume product is based on the network framework GCN[5- 8], and the traditional discrete convolution is applied to graph structure data. The convolution operator uses the position information in the graph to allow end-to-end learning of structured data, but its framework is only suitable for undirected graphs. Based on this, Schlichtkrullet al. [9] proposed theR-GCNmodel, which extendedGCNto the directed graph, and The model comprehensively considers all the information from neighboring entities in learning entity features. However, the learning results can't effectively measure the influence of neighboring entities on the current entities because the unified normalized constants are used in the learning process of neighboring entity information, and the correlation between entities is not considered.

Bilinear diagonal model DistMult [10] set each relationship type as diagonal matrix, and get the prediction score result through matrix multiplication and inner product operation on the feature vector space. This model greatly reduces the complexity of the model and is beneficial to deal with multi-relational data. butDistMultcan't handle the asymmetric relationship well, because the dot product calculation between real vectors is commutative, that is, if$(s, r \ \ , o)$ under the vector representation holds, then $(o, r, s)$ is bound to hold, but the proportion of asymmetric relations inKGis far more than that of symmetric relations. Therefore, the complex [11] model extends DistMultto the complex domain, and uses the characteristic that Hermite products between complex numbers are not commutative to model asymmetric relations. ComplExmodel captures the semantics of asymmetric relations and its prediction effect shows that the asymmetric relations in $KG$ are explicitly considered, which is beneficial to improve the reasoning effect. In addition, logical rules can also be used in KG reasoning [12-15].

Attention mechanism was put forward by Bahadanauand [16] in 2015, which simulates the brain signal processing mechanism unique to human vision, and can process variable-size input and pay attention to the most relevant parts to make a decision. Cheng Hua [17] and others pay attention to and weight the nodes in the topology diagram by using the attention mechanism, and strengthen the contribution of important nodes to the link prediction task. Petar [18] based on the attention mechanism, measures the correlation between adjacent entities, solves the node classification problem of relational graph data, and achieves good results in large-scale inductive data sets.

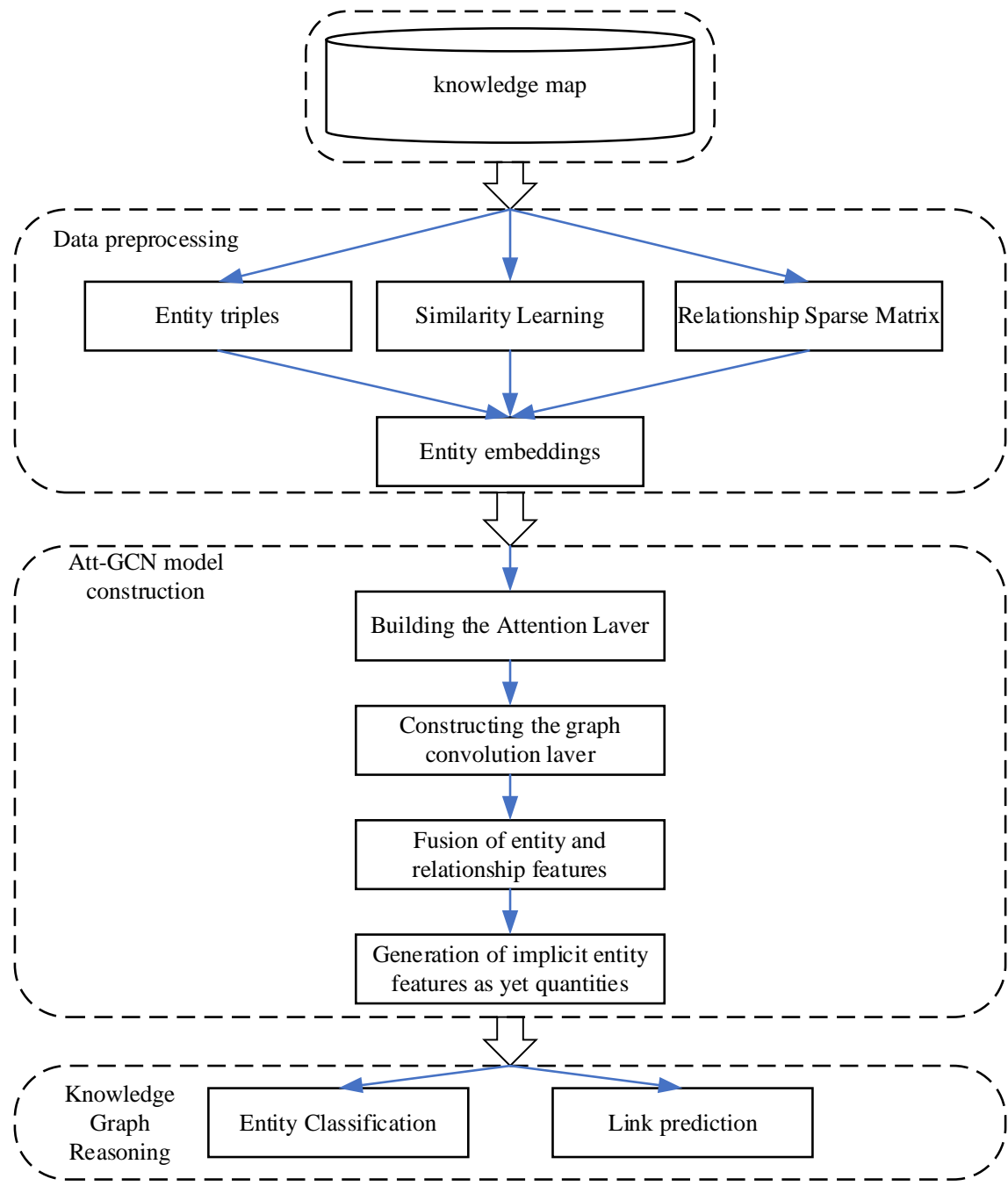

Figure 1 Model framework

Based on this, this paper puts forward a knowledge map reasoning model Att_GCN based on attention mechanism and graph convolution neural network, aiming at solving the problem that knowledge map reasoning based on neural network can not effectively measure the correlation between entities, and uses ComplEx as a scoring function to improve the accuracy of $KG$ reasoning by explicitly considering the asymmetric relationship in $KG$.

## 3 Knowledge Map Reasoning Model Based on Att_GCN

### 3.1 Pretreatment of Knowledge Map Data

$KG$ can be formally described as $G = \{V, E, \mathfrak{R}\}$, where $V$ represents the entity node set $v_i \in V$ in $KG$, $E$ represents the edge set $(v_i, r, v_j) \in E$, $\mathfrak{R}$ is the relation type set $r \in \mathfrak{R}$. Firstly, the entities and relationship types in KG are numbered to facilitate data processing and ensure their uniqueness. For example, triples (4512,52,546), where 4512 and 546 represent entity numbers and 52 represents relationship numbers, indicate that there is a relationship 52 between entities 4512 and 546. Because the representation of entity nodes in KG is discrete and unordered, the one-hot bag model is used to encode entities, and the entity embedding matrix G is generated in the order of entity numbers. Then, all the neighboring entities of any entity node $i$ are divided into two categories. Finally, the relationship sparse matrix $p$ of entities is generated, and the relationship sparse matrix of node $i$ is constructed according to the entity set information and relationship set information of node $i$, where the number of rows of the relationship sparse matrix represents all entities in $KG$, the number of columns represents the number of edges associated with the current entity, and the value represents the relationship type of the edge. Assuming that the triplet (4512,52,546) is the second side of the entity node 546, the value in the 4512th row and the second column of the forward sparse matrix $p'$ of the entity node 546 is 52.

### 3.2 Similarity Learning

A heterogeneous information network is a graph $\boldsymbol{G} = (\boldsymbol{V}, E, \boldsymbol{A}, X)$, where $\boldsymbol{V} \in \boldsymbol{R}^n$ represents the graph $\boldsymbol{G}$ of the nodes, the $n$ denotes the number of nodes; the $\boldsymbol{E}$ is a diagram $\boldsymbol{G}$ of the edge of the symbol; The adjacency matrix of the $\boldsymbol{A} \in \boldsymbol{R}^{n\times n}$ for $\boldsymbol{G}$; with regard to $V_{v_v,v_j} \in \boldsymbol{V}$, if $e_{ij} \in E$, 则 $A_{ij} = 1$, otherwise $A_{ij} = 0; X \in \boldsymbol{R}^{n\times m}$ denotes the attribute information of the node, the $m$ is the dimension of the attribute information.

GCN can be obtained by the convolution and aggregation operator [19] The low-dimensional embedding vectors of each node are learned from the neighboring nodes. The convolution operation for each layer is as follows.

$$\boldsymbol{H}_{i+1} = \delta\left(\hat{\boldsymbol{D}}^{-\frac{1}{2}}\hat{\boldsymbol{A}}\hat{\boldsymbol{D}}^{-\frac{1}{2}}\boldsymbol{H}_i W_i\right) \tag{1}$$

where, in the case of $\boldsymbol{H}_i$ is denoted by the first $i$ The output of layer i is also the output of layer $l$ of $l+1$ Layer of losers, the $\hat{\boldsymbol{A}} = \boldsymbol{A} + \boldsymbol{I}, \boldsymbol{I}$ denotes the deviation matrix; the $\hat{\boldsymbol{D}}$ The diagonal matrix represented by $\hat{D}_{ij} = \sum_{j=1}^{n} \hat{A}_{ij}; W_i$ is the number of $i$ the set of parameters of the layer; $\delta\delta(\cdot)$ represents an activation function.

During the model construction process, the $\boldsymbol{H}_{i+1}$ than $\boldsymbol{H}_i$ contains more high-level semantic information, yet ignores much of the low-level semantic information. For graphs, the low-level semantic information also plays an important role in the final embedding since it can reflect the local structural information of the graph.In addition, if the size of the input is large, the GCN which generates more parameters and is prone to overfitting.

Therefore, in this paper, the use of GCN Performing the similarity learning process, the $i-1$ Instead of splicing the output of the layer directly, the output of the layer is spliced through NNMF to the $\boldsymbol{H}_i (i = 0,1,2,\cdots,i-1)$ is decomposed into 2 low-dimensional matrices as follows.

$$\boldsymbol{H}_i = \boldsymbol{U}_i \boldsymbol{V}_i \text{ s.t. } \boldsymbol{U}_i \geqslant 0, \boldsymbol{V}_i \geqslant 0 \tag{2}$$

where $\boldsymbol{H}_i \in \boldsymbol{R}^{n\times L_i}, \boldsymbol{U}_i \in \boldsymbol{R}^{n\times d}, \boldsymbol{V}_i \in \boldsymbol{R}^{d\times L_i}, L_i$ is the number of i the output dimension of the layer, \ mathrm {i} $d$ is the output dimension of the nonnegative matrix factorization method, the $\boldsymbol{U}_i$ is $\boldsymbol{H}_i$ the representation vector of the lower dimensionality.Therefore, the first $i+1$ The output of the layer is represented as

$$\boldsymbol{H}_i' \leftarrow H_\tau \| U_0 \| \cdots \| U_{t-1} \tag{3}$$

where II denotes the splicing operation.

The low-dimensional representation vectors of the nodes are output from the GCN, and then the lowdimensional representation vectors are fed into a MultiLayer Perceptron (MLP) with a single hidden layer.The output from the MLP $\boldsymbol{Z}$ The similarity grouping, i.e.

$$\boldsymbol{Z} = \text{MLP}\ (\boldsymbol{H}_i') \tag{4}$$

where, in the case of $\boldsymbol{Z} \in \boldsymbol{R}^{n\times k}$.

## 3.3 Node Representation Learning

In this section, this paper carries out the representation learning for different nodes separately, and the realization process is as follows.

$$Q, \boldsymbol{K}, \boldsymbol{V} = \text{Embedding}(X)$$
$$\text{Attention}(Q, \boldsymbol{K}, \boldsymbol{V}) = \text{softmax}\left(\frac{Q\boldsymbol{K}^{\mathrm{T}}}{\sqrt{d_k}}\right)\boldsymbol{V} \quad (5)$$

$$\text{To } (Q, \boldsymbol{K}, \boldsymbol{V}) = (\text{head}_1, \cdots \text{ to } {}_h)\boldsymbol{W}^o$$

$$\text{head}_i = \text{Attention}\left(Q\, \mathrm{W}_i^Q, \boldsymbol{K}\mathrm{W}_i^{\boldsymbol{K}}, \boldsymbol{V}W_i^{\boldsymbol{V}}\right)$$

where $W_i^Q \in \boldsymbol{R}^{d_{\text{mode}} \times d_k}, W_i^K \in \boldsymbol{R}^{d_{\text{mode}} \times d_k}, W_i^V \in \boldsymbol{R}^{d_{\text{mode } 1} \times d_v}$, $W^o \in \boldsymbol{R}^{d_{\text{mod}} \times h d_v}$ represents a parameter vector.

The rest of the transformer layer is expressed as follows:

$$\begin{gathered} \boldsymbol{H}_2 = \text{FFN } (\boldsymbol{H}_1) \\ \tau = \text{LN } (\boldsymbol{H}_1 + \boldsymbol{H}_2) \\ \boldsymbol{H}_1 = \text{LN } (\text{Multihead}(Q, \boldsymbol{K}, \boldsymbol{V}) + X \end{gathered} \quad (6)$$

where LN $(\cdot)$ = Layer Normalization ( · ) denotes the layer normalization method, FFN $(\cdot)$ = Feed Forward Networks $(\cdot)$ represents a layer of feed-forward network.

Using Transformer to obtain context information vectors and combining with the similarity matrix obtained in Section 3.1, a graph attention network is employed to capture the interdependencies between nodes, thus realizing node representation learning.

Assuming that the graph contains $N$ nodes, and the feature vector of each node is $\boldsymbol{H}_i$, dimension $F$, as follows.

$$\boldsymbol{H} = \{\boldsymbol{H}_1, \boldsymbol{H}_2, \cdots, \boldsymbol{H}_N\} \quad (7)$$

A new feature vector $\boldsymbol{\delta}_i'$ I' can be obtained by linear transformation of the node feature vector $\boldsymbol{H}$, as shown below:

$$\begin{gathered} \boldsymbol{\delta}_i' = \boldsymbol{W}\boldsymbol{H}_i \\ \boldsymbol{\delta}' = \{\boldsymbol{\delta}_1', \boldsymbol{\delta}_2', \cdots, \boldsymbol{\delta}_N'\} \end{gathered} \quad (8)$$

where, the $\boldsymbol{W} \in \boldsymbol{R}^{F' \times F}$ is the matrix of linear transformations, the $F'$ represents the dimension of the transformation matrix.

Introducing the graph attention network, put the node $i, j$ The eigenvectors of $\boldsymbol{\delta}_i', \boldsymbol{\delta}_j'$ are spliced together, and then combined with a $2F'$ vector of dimension $\boldsymbol{a}$ Calculate the inner product.The activation function is a LeakyRelu function with the following formula: a

$$a_{ij} = \frac{\exp\left(\text{LeakyRelu}\left(\boldsymbol{a}^{\mathrm{T}}[W\boldsymbol{\delta}_i' \| W\boldsymbol{\delta}_j']\right)\right)}{\sum_{k \in N_i} \exp\left(\text{LeakyRelu}\left(\boldsymbol{a}^{\mathrm{T}}[W\boldsymbol{\delta}_i' \| W\boldsymbol{\delta}_j']\right)\right)} \quad (9)$$
$$\tilde{\boldsymbol{H}}_i' = \text{concat}\left(\sigma\left(\sum_{j \in N_i} a_{ij}^k \boldsymbol{W}^k \tau_j\right)\right)$$

where, the $\sigma$ denotes the activation function.

## 3.4 Att_GCN model construction

The construction of Att_GCN model mainly includes five steps: the design of attention layer, the design of graph convolution layer, feature fusion, entity classification and link prediction. The process of constructing Att_GCN model and feature fusion is shown in figure2.

### 3.4.1 Design of Attention Layer

This paper takes a single entity node $i$ as an example to illustrate the learning process of hidden feature vectors of each entity. According to the preprocessing process, the one-hot codes corresponding to the head entity set $N_{i_1}^r$ of node $i$ are $X = \{x_1, x_2, \cdots, x_N\}$, $x_j \in \mathbb{R}^M$, where $N$ is the number of entities in the set, $M$ is the dimension of each entity embedding vector, and $M$ is set to 300. In order to obtain enough expressive feature information in the attention layer and keep the same dimension in iterative training, the embedding vector $G$ input by the model is transformed linearly by sharing the weight $W \in \mathbb{R}^{M \times D}$, and then the correlation coefficient between node $j$ and node $i$ in the set is calculated by using the attention mechanism, and the nonlinear transformation is performed by LeakyReLU:

$$e_{ij} = \text{LeakyReL } U\left(a\left(\boldsymbol{W}\boldsymbol{x}_i, \boldsymbol{W}\boldsymbol{x}_j\right)\right) \quad (10)$$

Among them, attention mechanism $a$ is the inner product operation on $\mathbb{R}^D \times \mathbb{R}^D \rightarrow$, which measures the influence of any node $j$ on node $i$ in $N_{i_1}^r$ set. In order to make the $e_{ij}$ coefficient easy to compare on all entities in the set, the correlation coefficient obtained is normalized by using softmax mechanism.

$$\alpha_{ij} = \text{softmax}\left(e_{ij}\right) = \exp\left(e_{ij}\right) / \sum_{k \in N_{i_1}^r} \exp\left(e_{ik}\right) \quad (11)$$

Then $\alpha_{ij}$ is the degree of influence of neighboring node $j$ on $i$. Finally, the normalized weight coefficient $\alpha_{ij}$ is used to calculate the forward hidden state of node $i$:

$$h_{i_1}' = \text{relu}\left(\sum_{j \in N_{i_1}^r} \alpha_{ij} \boldsymbol{W} h_j\right) \quad (12)$$

Similarly, $N_{i_2}^r$ represents the set of all neighboring entities with entity node $i$ as the head entity, and the backward hidden state $h_i'$ of the corresponding node $i$ is obtained by using the same attention mechanism. The eigenvector learning of node $i$ comprehensively considers the effects of all its neighboring nodes on node $i$, which lays the foundation for the follow-up work.

### 3.4.2 Design of Volume Layer

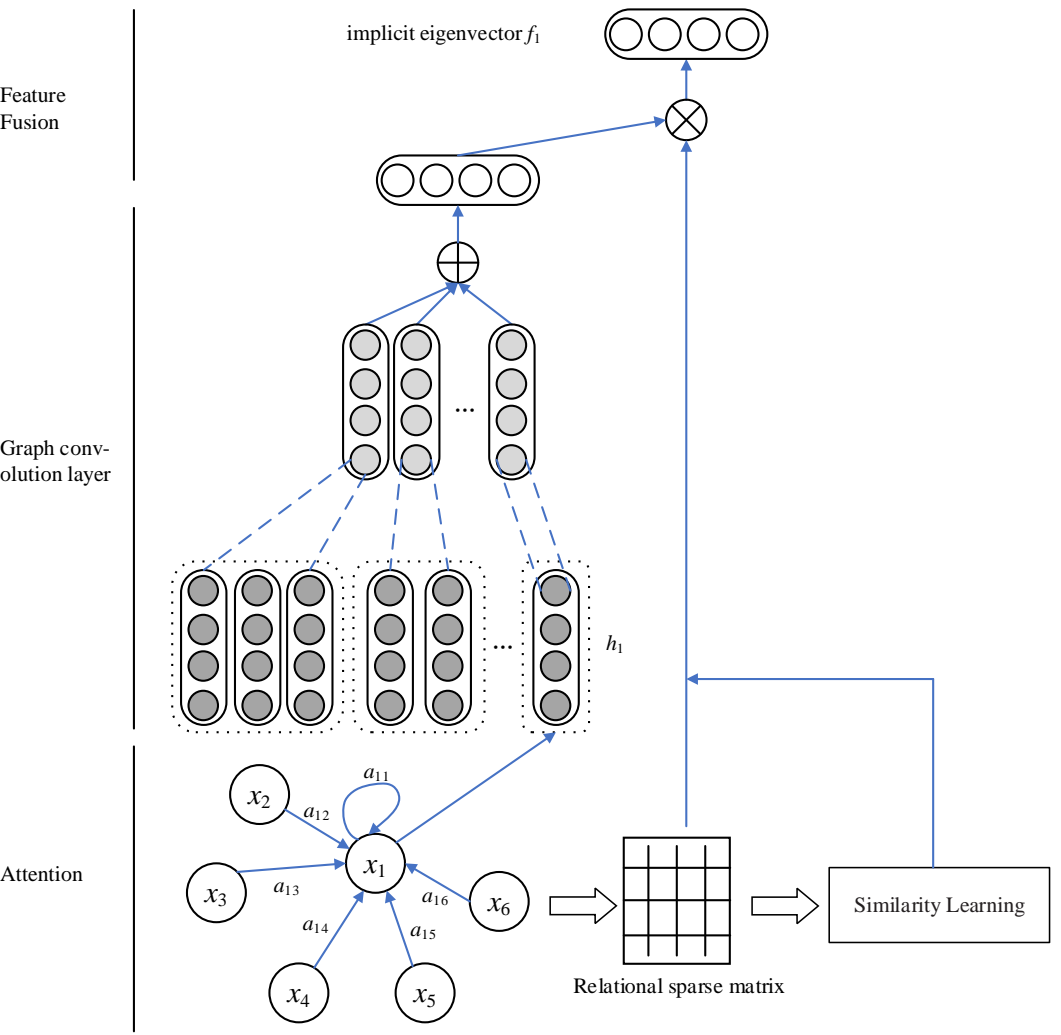

Fig. 2 Att_GCN model construction process

Graph convolution neural network realizes convolution operation on topological graph with the help of graph theory, and is used to extract spatial features. Using the parameter sharing technology of convolutional neural network, the parameters needed for feature learning can be reduced in the knowledge map inference with a large number of entity nodes, so that the parameters can be shared between rare relationships and frequent relationships, and the over-fitting problem of rare relationships can be effectively alleviated. Different from the ordinary graph convolution neural network, this paper uses a relationship-specific shared weight mechanism, that is, the determination of convolution kernel weight depends on the type and direction of edges. In the process of convolution operation, in order to ensure the correctness of neural transformation from $l$ layer to $l+1$ layer, a special relation transformation weight is set for each entity node to ensure the effective transmission of messages. In general, the convolution process from $l$ layer to $l+1$ layer is expressed as.

$$\boldsymbol{h}_i^{\prime(l+1)} = \sigma\left(\sum_{r\in\Re_j} \sum_{j\in N_{i_1}^r} \boldsymbol{W}_r^{(l)}\boldsymbol{h}_j^{(l)} + \boldsymbol{W}_0^{(l)}\boldsymbol{h}_i^{\prime(l)}\right) \quad (13)$$

Where $\boldsymbol{W}_r^{(l)}$ is the relation sharing weight of the entity set $N_{i_1}^r$ in the neighborhood of node. $h_i^{\prime(l+1)}$ represents the forward hidden state output of node $i$. The relationship sharing weight $W_r^{(l)}$ is determined by drawing lessons from the block decomposition model idea of R-GCN ([9]), so as to avoid the problem that the number of weight parameters increases rapidly due to the increase of the number of entities in $KG$, and fundamentally alleviate the over-fitting between rare relationships and oversize models. $W_r^{(l)}$ is directly defined by summing a set of low-dimensional matrices:

$$\boldsymbol{W}_r^{(l)} = \oplus_{b=1}^{B} \boldsymbol{Q}_{br}^{(l)} \quad (14)$$

Where $\oplus$ represents the splicing of matrices, so $W_r^{(l)}$ can also be represented as $\mathrm{diag}\left(Q_{1r}^{(l)}, Q_{2r}^{(l)}, \cdots, Q_{Br}^{(l)}\right)$ and $\boldsymbol{Q}_{br}^{(l)} \in \mathbb{R}^{\left(d^{\left(d^{(1)}/B\right)\times\left(d^{(l)}/B\right)}\right.}$. $\boldsymbol{W}_r^{(l)}$ can be regarded as a sparse constraint of weight matrices between different relationship types. Block decomposition can effectively reduce the number of parameters required by $KG$ and the over-fitting between rare relationships. By controlling the number of parameters of each weight matrix, the time complexity of model training can be reduced to some extent. Through the block decomposition model $W_r^{(l)}$, its parameters are grouped into sets of variables, which makes variables have higher coupling within groups than between groups.

Similarly, if the backward hidden state output $h_i^{\prime(l+1)}$ of node $i$ is learned by using the same relation sharing weight mechanism, the hidden state of node $i$ is $h_i^{(l+1)} = \boldsymbol{h}_i^{\prime(l+1)} + \boldsymbol{h}_i^{\prime\prime(l+1)}$.

In the process of model training, the learning process of hidden feature vectors of all entities is carried out in parallel, and the updating of each entity feature vector will affect the feature learning of all its adjacent entities. Therefore, the learning process of hidden feature vectors of entities is iterative until the state of each entity tends to be stable. Finally, the model can solve the problem of multi-step information transmission because the feature information of each entity is integrated with all the adjacent entity information.

#### 3.4.3 Integration of Entity and Relationship Features

Feature fusion is to fuse features from different sources and remove redundant information, so as to achieve the purpose of complementary advantages of multiple features. It is known that the hidden state of the neighborhood entity set $N_{i_1}^r$ of node $i$ is: $h^{\prime(l+1)} = \{h_1^{(l+1)}, h_2^{(l+1)}, \cdots, h_N^{(l+1)}\}, h_j^{(l+1)} \in \mathbb{R}^D$. Where $h_j^{(l+1)}$ represents the hidden state of node $j$. Because the relation sparse matrix of the relation set $r_1$ of node $i$ is $p_i^r$, the forward hidden feature vector of node $i$ is obtained by fusing the hidden state $\boldsymbol{h}^{\prime(l+1)}$ of the neighborhood entity set $N_{i_1}^r$ of node $i$ and the relation sparse matrix $\boldsymbol{p}_i^r$ of relation set $r_1$.

$$f_i' = \boldsymbol{p}_i' \otimes \boldsymbol{h}^{\prime(l+1)} \quad (15)$$

Among them, the corresponding backward hidden feature vector $f_{i\ i}$ can be obtained by the matrix multiplication of the sparse matrix $p_i^r$ and the hidden state $h^{\prime(l+1)}$, and the comprehensive hidden feature vector of node $i$ can be obtained by $f_i = f_i' + f''$.

### 3.5 Model Training

In order to evaluate the validity of the model, this paper tests it through two experiments: entity classification and link prediction.

The task of entity classification is to judge the semantic category of any given entity node $i$. This model trains all entity nodes by minimizing the cross entropy loss function:

$$L = -\sum_{i\in V} \sum_{k=1}^{K} t_{ik} \ln\left(\sigma\left(W^{D\times K} f_i\right)\right) \quad (16)$$

Among them, $t_{ik}$ represents the marker that node $i$ belongs to the K-th semantic category. When node $i$ belongs to the K-th semantic category, the value of $t_{ik}$ is 1, $W^{D\times K}$ is a simple linear transformation weight matrix, and $\sigma$ is a nonlinear transformation function ReLU.

The task of link prediction can be described as: giving a partial subset E of edge set e, designing a scoring function $f(s,r,o)$, and judging the possibility that any given edge $(s,r,o) \notin e$ belongs to E. In this paper, the complex [11] decomposition model based on complex domain space is used as the scoring function, and the embedded complex conjugate is used to deal with the asymmetric relationship in KG. By separating the real part and imaginary part of the entity feature vector and the relationship feature vector, the symmetrical relationship and anti-symmetrical relationship between entities can be accurately described, and the final scoring result is as follows:

$$\begin{aligned} f(s,r,o;\Theta) = \mathrm{Re}\,(< w_r, f_s, f_o >) = \\ \langle \mathrm{Re}\,(w_r), \mathrm{Re}\,(f_s), \mathrm{Re}\,(f_o) \rangle + \\ \langle \mathrm{Re}\,(w_r), \mathrm{Im}\,(f_s), \mathrm{Im}\,(f_o) \rangle + \\ \langle \mathrm{Im}\,(w_r), \mathrm{Re}\,(f_s), \mathrm{Im}\,(f_o) \rangle - \\ \langle \mathrm{Im}\,(w_r), \mathrm{Im}\,(f_s), \mathrm{Re}\,(f_o) \rangle \end{aligned} \quad (17)$$

$\mathrm{Re}\,(w_r)$ represents the symmetric real part of the relationship feature vector in $(s,r,o)$, and $\mathrm{Im}\,(w_r)$ represents the anti-symmetric imaginary part of the relationship feature vector in $(s,r,o)$. $\mathrm{Re}\,(f_s)$、$\mathrm{Im}\,(f_s)$ are the real and imaginary parts of the corresponding entity feature vectors in $(s,r,o)$.

During the experiment, the head or tail entities of some positive triples are randomly modified as negative samples for training, so that the ratio of positive and negative samples is 1:1, and the model results are optimized by cross entropy loss function:

$$L = -\frac{1}{2}\sum_{(s,r,o,y)\in\Gamma} ylb\ S(f(s,r,o;\Theta)) + \sum_{(s,r,o,y)\in\Gamma} (1-y)lb\,(1-S(f(s,r,o;\Theta))) \tag{18}$$

Where Γ is the set of all positive and negative samples, identified by $y$ , $y=1$ corresponds to the positive triplet, $y=0$ corresponds to the negative triplet, and $S$ is the logistic sigmoid function.

**Algorithm Att_GCN:**

input: Entity set $X$, relation set $R$, triple set $S$, maximum number of iterations $T$

output: updated entity set $X$.

/* Pretreatment*/

While $x_i \in X$

# Embedding $x_i$ into the vector space of $w$ dimension, entity set $X$, entity embedding dimension $w$, and shape is the number of entities * the dimension of each entity.

$x_i \leftarrow$ onehot_input ($X$,shape )

# Get the neighborhood entity matrix of node $i$, and find the embedding vector of each entity in $X$ through the index value $i$.

$N_i^r = t$ f.nn.embedding_lookup ($X$, index )

end while

while $r_i \in R$

# Generate the relationship vector of node $i$, and the weight parameters conform to the standard normal distribution, where the mean and standard deviation variance of the standard normal distribution, and the shape of the shared weight matrix is Shape.

$r_i \leftarrow$ make_tf_variable(mean, variance, shape, 'normal')

end while

while $s_i \in S, x_i \in X$

# Construct a sparse matrix of node $i$ relations, where indicators is a two-dimensional tensor (n,ndims ), $n$ is the number of non-zero elements, and ndims is the sparse matrix dimension; Values: the value corresponding to the position element indices by indicators; Dense_shape: the dimension of sparse matrix

$p'^r{}_i \leftarrow$ SparseTensor(indices, values, dense_shape) end while

/* Att_GCN model algorithm process */

for each $x_i \in X$

# Using the attention mechanism, the influence factor of neighbor entity $j$ on node $i$ is obtained, and $W$ is the weight parameter of the fully connected layer.

$\alpha_{ij} = \text{softmax}\left(\text{LeakyReL } U(Wx_i \cdot Wx_j)\right)$

# fuse the influence degree of neighborhood entities on node $i$ to get the hidden state of $i$.

$h_i' = \text{relu}\left(\sum_{j\in N_{i_1}^r} \alpha_{ij} W h_j\right)$

# Learn the neighborhood entity characteristics of node $i$ through graph convolution operation.

$h_i'^{(l+1)} = \sigma\left(\sum_{r\in\Re'}\sum_{j\in N_{i_1}^r} W_r^{(l)} h_j^{(l)} + W_0^{(l)} h_i'^{(l)}\right)$

# The hidden feature vector of node $i$ is obtained by fusing neighborhood entities and relationship features, and the neighborhood entity features of node $i$ are fused with sparse matrix information.

$f_i' = tf.\text{sparse_tensor_dense_matmul}\left(h_i'^{(l+1)}, p_i'\right)$

# Update the feature vector of node $i$, and fuse the forward information and backward information of node $i$ to get the whole information.

$x_i = f_i' + f''{}_i$

end for

## 4 Experimental Results and Analysis

### 4.1 Experimental Process of Entity Classification

The paper uses four RDF datasets: AIFB for predicting affiliations at AIFB research institute, MUTAG for cancer-related molecular data, BGS for rock type classification, and AM for cultural relic information. Each dataset involves classifying entity attributes as nodes, and Table 1 provides data statistics. Some relationships in the datasets have been removed, such as employs and Affiliation in AIFB, Isutagenics in Mutag, has Lithogenesis in BGS, and objectCategory and material in AM.

Table 1 Entity Classification Dataset

| Dataset | AIFB | MUTAG | BGS | AM |
|---|---|---|---|---|
| Number of entities | 8285 | 23644 | 333845 | 1666764 |
| Relation number | 45 | 23 | 103 | 133 |
| Edge number | 29043 | 74227 | 916199 | 5988321 |
| Number of labeled entities | 176 | 340 | 146 | 1000 |
| Class number | 4 | 2 | 2 | 11 |

Table 2 Comparison of Entity Classification Results

| Model | AIFB | MUTAG | BGS | AM |
|---|---|---|---|---|
| R-GCN | 95.83 | 75.23 | 83.10 | 89.29 |
| Att-GCN | 97.98 | 77.23 | 85.41 | 90.79 |

Parameter setting: 500-dimensional entity training dimension is used in this paper, which is consistent with the R-GCN model to ensure the fairness of the experiment. At the same time, through the experimental comparison of different entity embedding dimensions, it is known that the training effect is the best when the entity dimension is 500. In order to speed up the training and avoid over-fitting in the training process, $L2$ regularization is used to constrain the relationship sharing weight parameters, because $L2$ regularization assumes that the prior distribution of parameters is Gaussian distribution, which can ensure the stability of the model and the value of parameters will not be too large or too small, and the penalty value $L2$ ranges from $\{0,0.0005,0.001\}$. And using the dropout strategy, the setting range is $d=\{0.3,0.4,0.5,0.6,0.7\}$; The size parameters of each batch of training data in the iterative process are $\{50, 100,150,200\}$; The full batch gradient descent technique is used in the model training process. All training only optimizes the training set with super parameters. Based on the performance of the verification set, $L2 = 0.0005$, attention layer $d = 0.6$, volume layer $d = 0.4$, and the size of training data sent in each batch is 50. The results of the comparison model are directly obtained from the paper, and the evaluation results are shown in Table 2.

Experimental analysis: As shown in Table 2, for the four data sets, the classification results of Att_GCN are all improved by about 2% compared with the improved R-GCN model. The experimental results show that learning messages from neighboring entities in different degrees can pay more attention to neighboring entities with great influence and weaken some useless information, so that the learned entity feature vectors are more in line with the actual situation.

Table 3 The results of the comparison experiment Hits@10 of link prediction parameters

| Iterations | Entity dimension | Number of layers of graph volume product | | |
|---|---|---|---|---|
| | | 1 | 2 | 3 |
| 3000 | 300 | 0.331 | 0.387 | 0.375 |
| | 400 | 0.334 | 0.388 | 0.381 |
| | 500 | 0.343 | 0.389 | 0.387 |
| 4000 | 300 | 0.342 | 0.391 | 0.383 |
| | 400 | 0.345 | 0.389 | 0.389 |
| | 500 | 0.357 | 0.398 | 0.996 |
| 5000 | 300 | 0.349 | 0.392 | 0.389 |
| | 400 | 0.353 | 0.398 | 0.392 |
| | 500 | 0.363 | 0.412 | 0.403 |
| 6000 | 400 | 0.362 | 0.410 | 0.399 |
| | 500 | 0.377 | 0.433 | 0.409 |
| | 600 | 0.378 | 0.427 | 0.412 |
| 7000 | 400 | 0.361 | 0.402 | 0.401 |
| | 500 | 0.374 | 0.411 | 0.413 |
| | 600 | 0.379 | 0.419 | 0.417 |

Table 4 Comparison of Link Prediction Results

| model | MRR | | Hits@ | | |
|---|---|---|---|---|---|
| | Raw | Filtered | 1 | 3 | 10 |
| DistMult | 0.100 | 0.191 | 0.106 | 0.207 | 0.376 |
| ComplEx | 0.109 | 0.201 | 0.112 | 0.213 | 0.388 |
| R-GCN | 0.158 | 0.248 | 0.151 | 0.264 | 0.417 |
| Att_GCN | 0.189 | 0.272 | 0.172 | 0.294 | 0.450 |

### 4.2 Experimental Process of Link Prediction

Dataset: The challenging FB15K-237 dataset [19] is selected as the evaluation object. Because there are a large number of inverse triples on WN18k [19] and FB15K [19] data sets, the simple linear transformation model Link-Feat [19] based on rules can get good prediction results by using the inverse triples information. FB15K-237 data set is based on FB15K, and all inverse triples are removed. Through this data set, the performance of the model can be better evaluated. FB15K-237 contains 14541 entities and 237 relationship types.

Parameter setting: The entity training dimension selected in the experiment is 500 dimensions, and $L2$ regularization is used to constrain the scoring function. The penalty value $L2$ ranges from $\{0,0.005,0.01,0.015,0.02\}$ , the learning rate is $\{0,0.001,0.01\}$, and the number of layers of stacked graphs is $c = \{1,2,3\}$. By verifying the experimental results, $L2 = 0.01$, attention layer $d = 0.6$, volume layer $c = 2$, volume layer $d = 0.5$, and the number of iterations is 6000. The parameter setting of the number of training iterations, the physical training dimension and the number of layers of the graph volume during the experiment is shown in Table 3. Experimental analysis: From Table 4, it can be seen that the experimental results of Att_GCN are higher than those of DisMult model, directly improved ComplEx model and R-GCN model.

## 5 Conclusion

Since each entity in KG needs to learn its corresponding neighborhood entity set information, this method needs to perform convolution operations for each entity for many times, so it is difficult to greatly reduce the time complexity. In the future, it is considered that the head entity and relationship should be regarded as a whole, and all entities in the data set should be regarded as tail entities for link prediction, so as to achieve the purpose of reducing the number of convolution operations. Secondly, the object of reasoning is closed KG, that is, the relationship and entity types of KG are fixed, and new entities and relationships cannot be added. How to apply the deep learning method to the open field for knowledge reasoning and automatic discovery of new relationships needs further study.